\documentclass[letterpaper,reprint, aps, prl, superscriptaddress]{revtex4-1}
\usepackage[T2A,T1]{fontenc}
\usepackage[latin9]{inputenc}
\usepackage{textcomp}
\usepackage{amsmath}
\usepackage{amssymb, bm}
\usepackage{graphicx}
\usepackage{placeins}


\begin{document}

\title{Quantum Critical Phenomena in a Compressible Displacive Ferroelectric}

\author{Matthew J Coak}
\affiliation{Cavendish Laboratory, Cambridge University, J.J. Thomson Ave, Cambridge CB3 0HE, UK}

\author{Charles R S Haines} 
\affiliation{Cavendish Laboratory, Cambridge University, J.J. Thomson Ave, Cambridge CB3 0HE, UK}

\author{Cheng Liu}
\affiliation{Cavendish Laboratory, Cambridge University, J.J. Thomson Ave, Cambridge CB3 0HE, UK}

\author{Stephen E Rowley}
\affiliation{Cavendish Laboratory, Cambridge University, J.J. Thomson Ave, Cambridge CB3 0HE, UK}
\affiliation{Centro Brasileiro de Pesquisas Fisicas, Rua Dr. Xavier Sigaud 150, Rio de Janeiro 22290-180, Brazil}

\author{Gilbert G Lonzarich}
\affiliation{Cavendish Laboratory, Cambridge University, J.J. Thomson Ave, Cambridge CB3 0HE, UK}

\author{Siddharth S Saxena}
\affiliation{Cavendish Laboratory, Cambridge University, J.J. Thomson Ave, Cambridge CB3 0HE, UK}
\affiliation{National University of Science and Technology \textquotedblleft MISiS\textquotedblright, Leninsky Prospekt 4, Moscow 119049, Russia}

\date{\today}

\begin{abstract}
The dielectric and magnetic polarizations of quantum paraelectrics and paramagnetic materials have in many cases been found to initially increase with increasing thermal disorder and hence exhibit peaks as a function of temperature. A quantitative description of these examples of 'order-by-disorder' phenomenona has remained elusive in nearly ferromagnetic metals and in dielectrics on the border of displacive ferroelectric transitions.  Here we present an experimental study of the evolution of the dielectric susceptibility peak as a function of pressure in the nearly ferroelectric material, strontium titanate, which reveals that the peak position collapses towards absolute zero as the ferroelectric quantum critical point is approached.  We show that this behaviour can be described in detail without the use of adjustable parameters in terms of the Larkin-Khmelnitskii-Shneerson-Rechester (LKSR) theory, first introduced nearly 50 years ago, of the hybridization of polar and acoustic modes in quantum paraelectrics, in contrast to alternative models that have been proposed.  Our study allows us to construct for the first time a detailed temperature-pressure phase diagram of a material on the border of a ferroelectric quantum critical point comprising ferroelectric, quantum critical paraelectric and hybridized polar-acoustic regimes.  Furthermore, at the lowest temperatures, below the susceptibility maximum, we observe a new regime characterized by a linear temperature dependence of the inverse susceptibility that differs sharply from the quartic temperature dependence predicted by the LKSR theory.  We find that this non-LKSR low temperature regime cannot be accounted for in terms of any detailed model reported in the literature, and its interpretation poses a new empirical and conceptual challenge. 
\end{abstract}

\maketitle

The study of quantum phase transitions and quantum critical systems has led to the discovery of novel phases of matter and the introduction of novel conceptual frameworks for the description of emergent phenomena~\citep{Sachdev2011}. A quantum phase transition reached by varying a tuning parameter such as lattice density or electronic band filling fraction is imagined to separate two or more low-temperature states with qualitatively different types of order. An example is a transition from a magnetically polarized to a paramagnetic state in a metal. In the Kondo lattice model, for instance, at suffficiently low temperature the paramagnetic state is described as a Fermi liquid in which the elementary excitations arise from the hybridization of conduction electron states and well localized f-electron orbitals.

Another example involves a transition from a displacive ferroelectric state to an unpolarized or quantum paraelectric state in polar materials such as the perovskite oxides \citep{Rechester1971,Khmelnitskii1971,Khmelnitskii1973,Hoechli1979,Bednorz1984,Kvyatkovskii2001,Roussev2003,Venturini2004,Wu2006,Das2009,Palova2009,RowleyThesis,Rowley2010,Rowley2014,Rowley2015,Rowley2014a,Horiuchi2015,Rowley2016,Chandra2017a,Rischau2017a}.
In contrast to the case of the magnetic metals the nature of the unpolarized state in incipient ferroelectrics remains in some respects an enigma, especially in the low temperature regime. In the simplest model, the quantum paraelectric state is characterized by an activated form of the temperature dependence of the inverse dielectric susceptibility in which the activation temperature scale vanishes at a continuous quantum phase transition, i.e. at a quantum critical point. However, this picture has proved to be insufficient and in particular is contradicted by the observation of an anomalous temperature dependence - including a mysterious minimum - in the inverse susceptibility of SrTiO$_3$ and related incipient displacive ferroelectrics at low temperatures~\citep{Rowley2014,Rowley2015,Rowley2016,Coak_Thesis_2017}, which theoretical works have attempted to describe \citep{Khmelnitskii1971,Palova2009,Rowley2014}. 

The identification of the nature of the quantum paraelectric states in such materials has been a key objective of the present study. This is a part of a more general goal to characterize and understand the temperature-quantum tuning parameter phase diagram of quantum critical ferroelectrics. 

The absence of free charge carriers (in undoped samples) was expected to lead to a simpler phase diagram than that observed near to quantum critical points in metals, in which quantum critical phenomena are in many interesting cases masked by the emergence of intervening phases. These include unconventional superconductivity and exotic textured phases, which are of great interest but stand in the way of understanding quantum critical behaviours in their simplest forms over wide ranges down to very low temperatures. 

To characterize the temperature-quantum tuning parameter phase diagram in close detail and obtain a deeper understanding of the quantum paraelectric state we have carried out measurements of the dielectric response over a wide range in temperature and pressure with high precision. In particular, the identification of the low temperature behaviour of the relative dielectric constant, $\varepsilon_{r}$, or dielectric susceptibility, $\chi=\varepsilon_{r}-1$, has benefited from measurements of relative changes of $\chi$ as small as a few parts per billion. We first mention briefly the results of some relevant previous studies of our chosen material and then present and discuss our new findings. 

SrTiO$_3$ is a well-studied incipient displacive ferroelectric~\citep{Lines1977}, widely used as a dielectric in deposition techniques and thin-film interface devices~\citep{Atkinson2017}, as well as recently in high-precision thermometry \citep{Tinsman2016}, and is remarkable for having an extremely high dielectric susceptibility at low temperatures. At high temperatures a good fit to the classically predicted Curie-Weiss form of the dielectric susceptibility is observed, with an extrapolated Curie temperature around 35~K \citep{Muller1991}, but this temperature dependence changes below approximately 50~K in the quantum critical regime and no ferroelectric ordering occurs down to the lowest temperatures measured. In addition, substitution of oxygen-16 for the oxygen-18 isotope results in the material becoming ferroelectric, and varying the level of isotope substitution or applying pressure (to samples with suffficiently high oxygen-18 concentrations) tunes the Curie temperature $T_{c}$ to zero~\citep{Itoh2000a,Wang2000}. Uniaxial tensile strain applied to SrTiO$_3$ again causes it to become ferroelectric and suggests a small negative critical pressure of magnitude of the order of one kbar~\citep{Uwe1976}. Measurements of the dielectric susceptibility under pressure~\citep{Hegenbarth1967,Frenzel1974,Pietrass1972} show a drastic suppression of the low temperature dielectric response as pressure is increased, matching the trend seen in the oxygen isotope doping studies which see a maximum in the size of $\chi$ at a substitution level of 36\%, the same point where $T_{c}$ tends to zero temperature. At this quantum critical point the frequency of the polar transverse optical phonon mode responsible for the ferroelectric ordering approaches zero at the Brillouin zone centre~\citep{Uwe1976}. Recent work~\citep{coak2018b} has shown that the magnitude of the dielectric loss peak at approximately 10 K, associated with quantum critical effects~\citep{Viana1994}, is linked to the quantum critical point in agreement with results from oxygen-18 substituted SrTiO$_3$~\citep{Venturini2004}. An open question in the field remains as to the quantum phase transition empirically not becoming first-order as temperature is lowered and lifting the quantum criticality~\citep{Chandra2018}. Although it is reasonable to expect that a ferroelectric transition such as this would turn first order, there is overwhelming evidence that the system is indeed quantum critical. Further work in the field is needed to advance understanding on this apparent contradiction.

These and related studies, including those on superconductivity in doped SrTiO$_3$ \citep{Schooley1964,Marel2011,Scott2012,Lopez-Bezanilla2015,Rowley2018,Lin2013,Ueno2008,Fernandes2013,Klimin2014,Lin2014,Edge2015,Ruhman2016,Chubukov2016,Gorkov2016,Rischau2017a,Stucky2016},
shed light on the likely broad features of the temperature-quantum tuning parameter phase diagram of SrTiO$_3$. We now turn to our present findings that allow us to construct the first detailed phase diagram of this kind, with hydrostatic pressure, that preserves the high degree of homogeneity of the starting material, as the chosen quantum tuning parameter. Importantly, our results enable us to identify the physical nature of the quantum paraelectric state at pressures above the critical pressure of the ferroelectric quantum critical point at low temperatures, and in particular below the ubiquitous peak in the dielectric susceptibility.

\section*{Results}

\begin{figure}
\centering
\includegraphics[width=1\columnwidth]{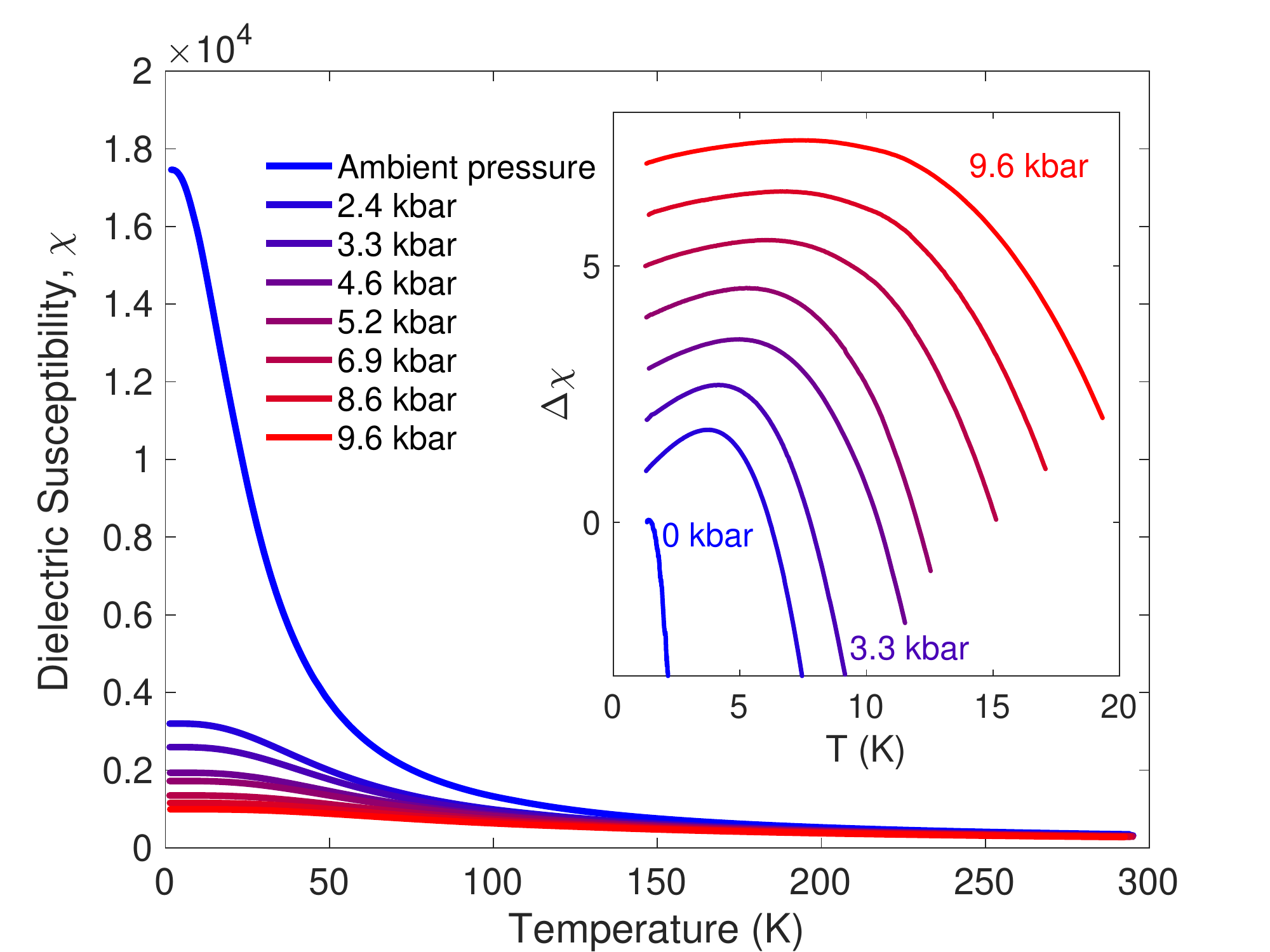}
\includegraphics[width=1\columnwidth]{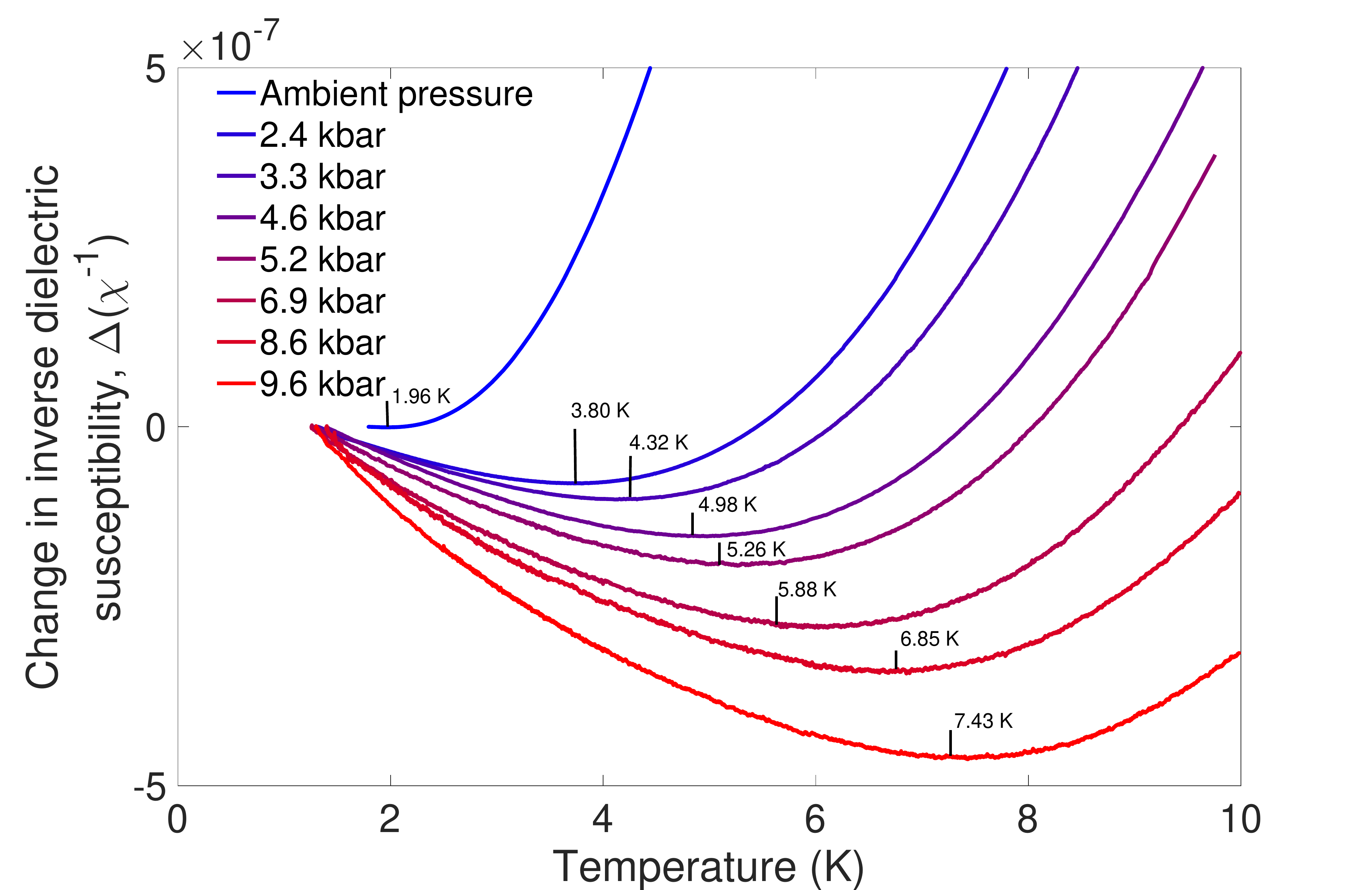}
\caption{Upper plot - the dielectric susceptibility $\chi$ of SrTiO$_3$ plotted against temperature for applied pressures ranging from 0 (blue) to 9.6 (red) kbar. The magnitude of the dielectric susceptibility can be seen to be continuously reduced by the application of pressure. The inset shows the change in the low temperature values of $\chi$ from their lowest-temperature values; curves are offset for clarity. Importantly, $\chi$ initially rises with temperature and exhibits a peak that increases in position and magnitude with increasing pressure. The lower plot shows the change in the inverse of $\chi$ from its lowest-temperature values for each pressure (typically 1.6 K), where the feature is now a clearly resolved minimum.}
\label{fig:STO dielectric plot}
\end{figure}

Fig.~\ref{fig:STO dielectric plot} shows measurements of the dielectric susceptibility $\chi=\epsilon_r-1$ of SrTiO$_3$ at ambient pressure and at increasing applied pressures. The ambient pressure data match the results of earlier work~\citep{Mueller1979,Rowley2014} wherein the inverse susceptibility is linear at high temperatures matching the expected Curie-Weiss behaviour, before crossing over to a quadratic power law dependence at lower temperatures attributed to quantum critical fluctuations. The low temperature dielectric susceptibility reaches a maximum at approximately 2~K with a value of around 20,000 before falling at even lower temperatures. Observed as a minimum in the inverse susceptibility, this effect is resolved here in much clearer detail than in earlier studies \citep{Wang2000,Rowley2014} and crucially is investigated as a function of pressure. In the lower plot of Fig. \ref{fig:STO dielectric plot} this minimum is seen to increase in depth with increasing pressure and its position, marked with arrows, moves up in temperature.

\begin{figure}
\centering
\includegraphics[width=1\columnwidth]{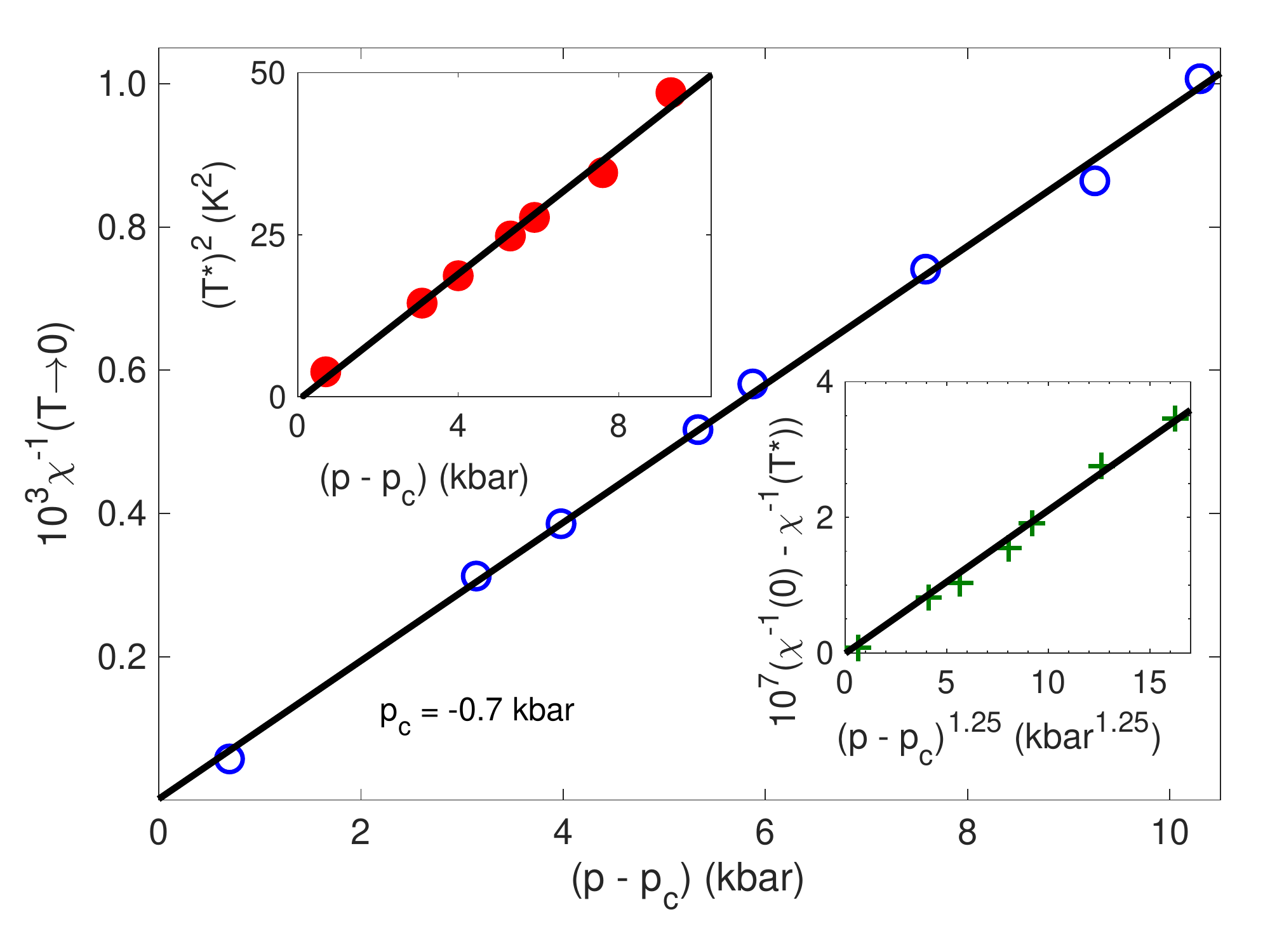}
\caption{The low temperature inverse dielectric susceptibility $\chi^{-1}(0)$ = $\chi^{-1}(T\rightarrow0)$ as a function of applied pressure. We see that $\chi^{-1}(0)$ varies linearly with pressure and vanishes at the extrapolated critical pressure, $p_c$, of -0.7(1)~kbar, defining the ferroelectric quantum critical point. The lower and upper insets show, respectively, the temperature dependences of the position of the minimum, $T^*$, and of the depth of the minimum, $\Delta\chi^{-1}(T^*)$. We see that the square of $T^*$ is proportional to pressure and hence also to $\chi^{-1}(0)$. This is characteristic of the model of coupled polar and non-polar modes, i.e. the LKSR model, as described in the text. The solid lines in all three plots are guides to the eye.}
\label{fig:STO linear plots}
\end{figure}

Key features of the susceptibility are brought out in Fig. \ref{fig:STO linear plots}, which shows the pressure dependences of the $T\rightarrow0$~K inverse susceptibility $\chi^{-1}(0)$, (main plot), the square of the position of the minimum $T^*$ (upper inset) and the depth of the minimum $\Delta\chi^{-1}(T^*) = \chi^{-1}(0)-\chi^{-1}(T^*)$ (lower inset). All three curves extrapolate to zero at the same critical pressure, $p_c = -0.7(1)$~kbar, i.e., at the ferroelectric quantum critical point, as suggested above. We see that $\chi^{-1}(0)$ varies linearly with $(p-p_{c})$, $T^*$ varies as the square root of $(p-p_{c})$ and $\Delta\chi^{-1}(T^*)$ varies as $(p-p_{c})$ to a power slightly greater than unity. 

\begin{figure*}
\centering
\includegraphics[width=1.6\columnwidth]{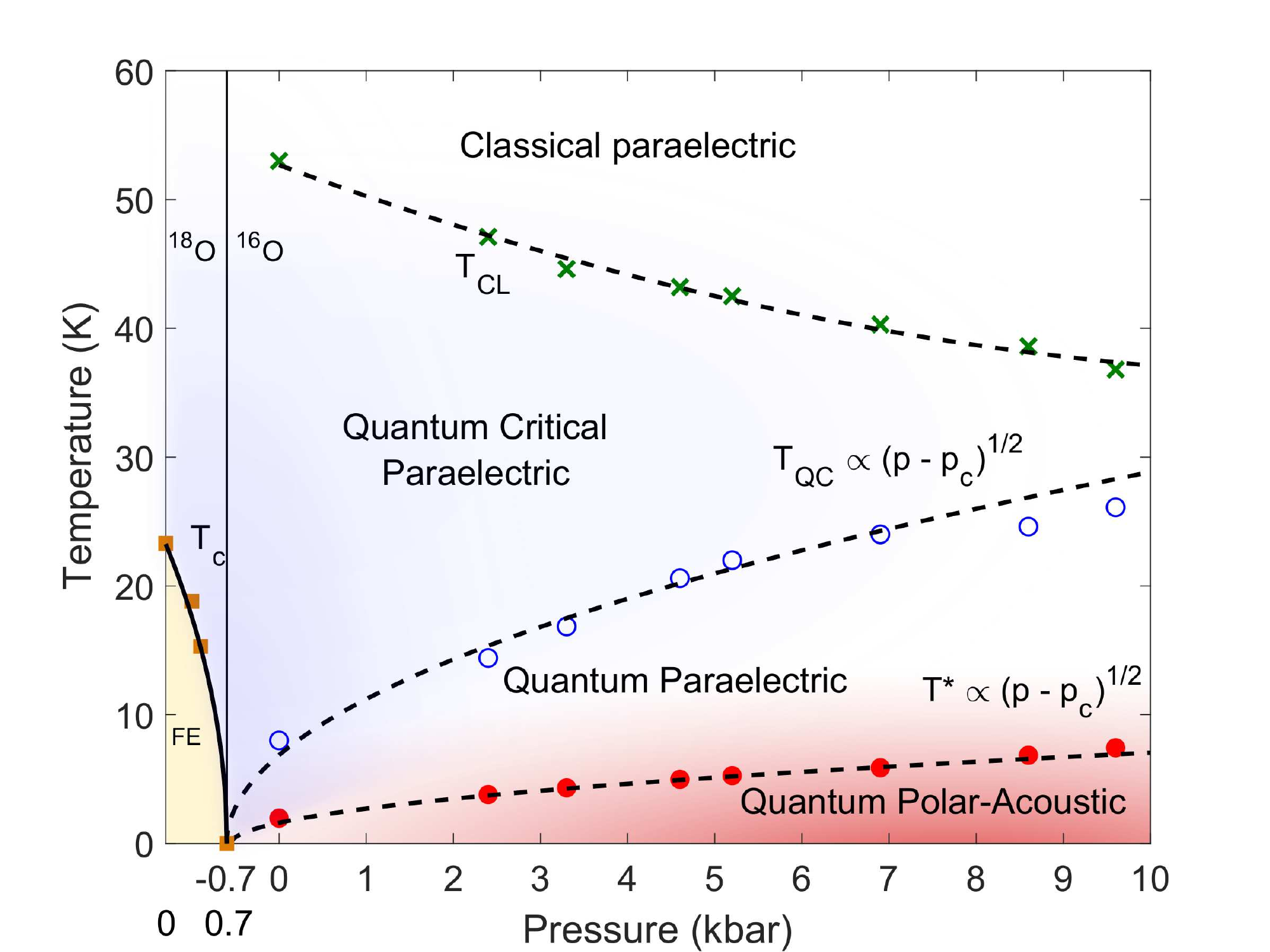}
\caption{Phase diagram for SrTi$^{16}$O$_3$ from -0.7 to 10~kbar (right) and SrTi$^{18}$O$_3$ from 0 to 0.7~kbar (left), overlayed to match up the positions of proposed QCPs. Closed circles give the positions of the low temperature minimum, $T^*$, in $\chi^{-1}$ and open circles the crossover temperature, $T_{QC}$ from the quantum paraelectric to quantum critical regimes. Dashed lines give fits of $(p-p_c)^{1/2}$ behaviour to both. Crosses show the crossover temperature, $T_{CL}$, between quantum critical and classical Curie-Weiss behaviour with a dashed guide to the eye. Squares and solid line show the ferroelectric Curie temperature, $T_{c}$, of SrTi$^{18}$O$_3$ taken from \citep{Wang2000}.}
\label{fig:FullPhaseDiagram}
\end{figure*}

The variation of $\chi^{-1}(T)$ above a scale $T_{QC} >T^*$, which vanishes along with $T^*$ at $p_{c}$, is found to be quadratic, $T^{2}$, up to another scale $T_{CL}$, and is linear in the classical regime above $T_{CL}$ (SI Appendix and Fig. \ref{fig:FullPhaseDiagram}). Pressure narrows the temperature window of the {$T^{2}$} quantum critical regime between $T_{QC}$ and $T_{CL}$ while widening that below $T_{QC}$, including the interesting regime below $T^*$. Please see the SI Appendix and references therein for the error analysis and a full discussion of the fitting processes used in defining $T_{QC}$ and $T_{CL}$. Combining the data for the pressure-dependent temperatures of the low-temperature minimum, $T^*$, (Fig. \ref{fig:STO dielectric plot} and Fig. \ref{fig:STO linear plots} upper inset) and the crossover temperatures from quantum paraelectric to quantum critical and from quantum critical to classical regimes, $T_{QC}$ and $T_{CL}$, respectively, with previous data on SrTi$^{18}$O$_3$ \citep{Wang2000} that yields the Curie (critical) temperature, $T_{c}$, under pressure allows a full mapping of the temperature-pressure phase diagram of SrTiO$_3$, which is shown in Fig. \ref{fig:FullPhaseDiagram}. The single ferroelectric quantum critical point at $p_{c}$ is shown to be the origin of both the $T^{2}$ region of quantum critical behaviour, and seemingly the energy scale of the minimum feature $T^*$ - suggesting this effect to emanate from the QCP.

\section*{Discussion}

The main features of this phase diagram are consistent with the predictions of a 3-D self-consistent Gaussian mean field model, also known as the self-consistent phonon model (see e.g. \citep{Rowley2014} and references therein), which assumes that $\chi^{-1}(0)$ is an analytic function of $(p-p_{c})$ (in analogy to the assumption of analyticity in the Landau theory of phase transitions at finite temperatures) and that the temperature dependence $\chi^{-1}(T)$ is due to the thermal excitation of polar transverse optical modes whose gap, $\Delta$, vanishes at the quantum critical point. The contribution of each mode depends on the inverse of their wavevector, so that in three dimensions at the quantum critical point one expects a contribution to $\chi^{-1}(T)$ of the form $(1/T)T^{3}$, or $T^{2}$, far below the relevant Debye temperature, $T_{D}$, and of the form $T$ above a temperature, $T_{CL}$, calculated numerically to be a sizeable fraction of $T_{D}$ \citep{Rowley2014}. Away from the quantum critical point where $\Delta$ is finite the $T^{2}$ regime is cut off below a scale $T_{QC}$ where the temperature dependence becomes exponentially weak as expected for activated phenomena. Since $\Delta^{2}$ is expected to be proportional to $\chi^{-1}(0)$, which is proportional to $(p-p_{c})$, we expect $T_{QC}$ to be proportional to the square root of $(p-p_{c})$, which is in keeping with observation (Fig. \ref{fig:FullPhaseDiagram} and SI Appendix). Similar considerations lead us to expect $T_{c}$ to also be proportional to the square root of $(p-p_{c})$, which is consistent with previous studies in SrTi$^{18}$O$_3$. As shown previously for ambient pressure measurements, the self-consistent phonon model provides not only a qualitative but also quantitative understanding of the above behaviour in terms of independently measured model parameters.

However, the self-consistent phonon model in its simplest form fails to account for the low temperature behaviour presented here for $T<T^*$, which suggests that the quantum paraelectric state at low $T$ is very different from the traditionally accepted gapped state with activated behaviour, e.g. as described by the Barrett theory \citep{Barrett1952}. In the remainder of the paper we consider alternative possible descriptions of this state and attempt to clarify its physical nature. 

We discuss first the role of the coupling of the electric polarization with the non-polar lattice vibrations or acoustic phonons not included in the above self-consistent phonon model. As shown previously \citep{Khmelnitskii1971,Palova2009,Rowley2014} this coupling can account for the existence of a minimum of the inverse susceptibility with values of $T^*$ and depth $\Delta\chi^{-1}(T^*)$ that are consistent with zero-temperature model parameters inferred from other measurements. Extending measurements to include the effect of pressure tuning, however, sheds light on a particularly distinctive prediction of the model, namely that the square of $T^*$ should vary linearly with $\chi^{-1}(0)$ and hence vanish at the ferroelectric quantum critical point. This self consistent phonon theory including polarization-acoustic phonon couplings is referred to here as the Larkin-Khemelnitskii-Shneerson-Rechester (LKSR) theory \citep{larkin1969a,larkin1969b,Khmelnitskii1971,Palova2009,Rowley2014}.

\begin{figure}
\centering
\includegraphics[width=1\columnwidth]{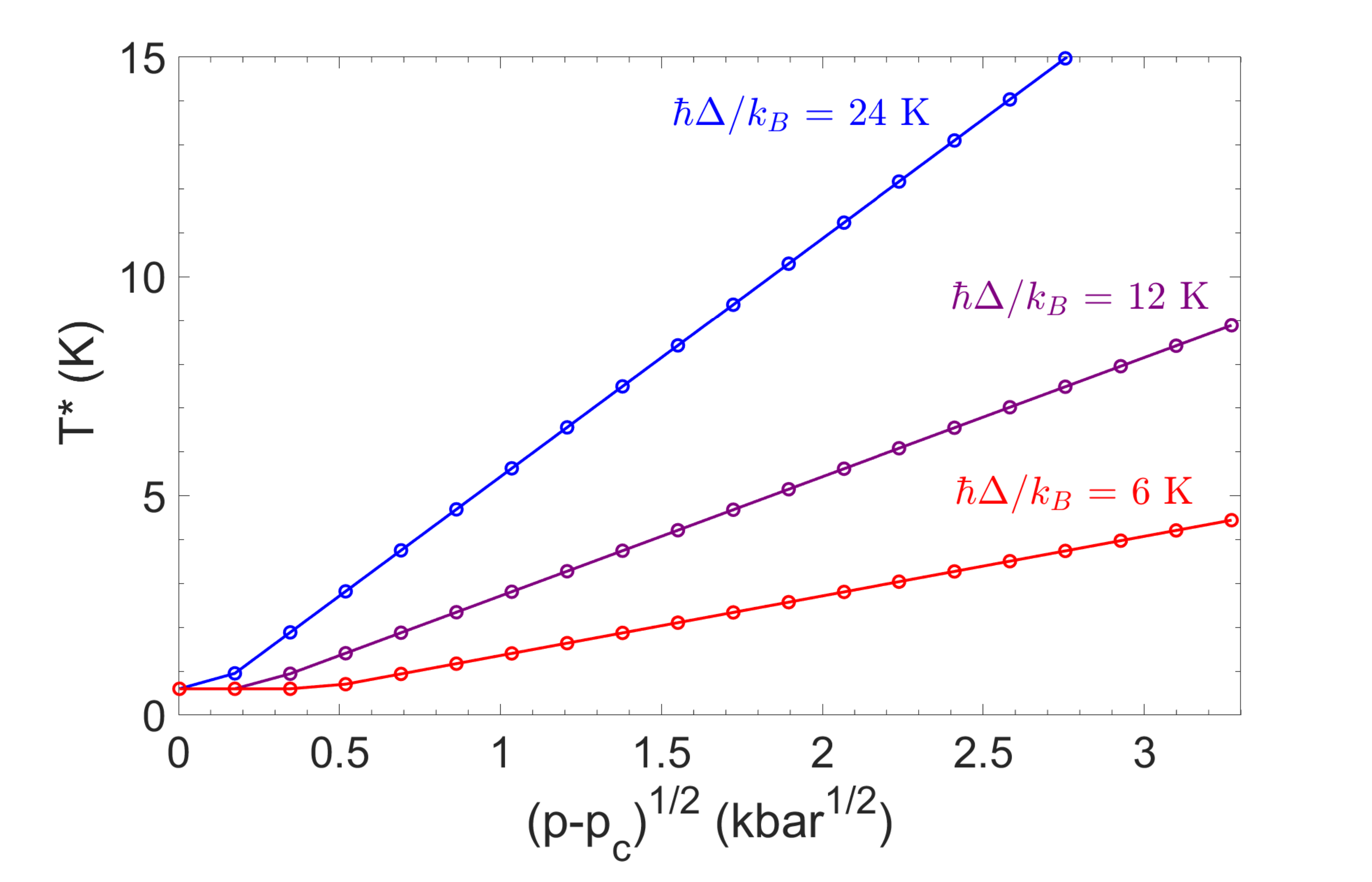}
\caption{Calculated pressure dependence of the temperature $T^*$  of the 		minimum of the inverse susceptibility. Predictions of the self-consistent phonon model including the electrostrictive
coupling (Equations 1-5 in the SI Appendix) for $T^*$ vs $(p-p_{c})^{1/2}$ of the inverse susceptibility vs temperature, for three values of the low-temperature and zero-pressure gap $\hbar\Delta/k_{B}=$ 24~K (blue), 12~K (purple) and 6~K (red) (see SI Appendix for the definition and determinations of the model parameters). The square of $T^*$ is proportional to the pressure change measured from the ferroelectric quantum critical point at $p_c$, in agreement with observation (Fig. \ref{fig:STO linear plots}).}
\label{fig:Calc_temp_var_comp}
\end{figure}

This prediction is strikingly supported by the data presented in the main plot and upper inset of Fig. \ref{fig:STO linear plots}, which show that both $(T^*)^2$ and $\chi^{-1}(0)$ vary linearly with $(p-p_{c})$ and hence are proportional to each other (see SI Appendix, Fig S5). Moreover, as shown in the SI Appendix, the absolute value of the slope of $(T^*)^2$ vs $p - p_{c}$ or equivalently $\chi^{-1}(0)$ is consistent in order of magnitude with independently measured model parameters - in particular the results of the calculations for $T^*$ vs. the square root of $(p-p_{c})$ are shown in Fig. \ref{fig:Calc_temp_var_comp}. Thus, at the critical pressure, $p_{c}$, $\chi^{-1}(T)$ has no minimum and is predicted to vary as the square of the temperature down to the lowest temperatures. For our model parameters the transition to the ferroelectric state is expected to be essentially continuous at low temperatures, despite the polarization-acoustic phonons coupling (electrostriction) that is often expected to lead to first order transitions. This prediction appears to be in keeping with measurements to date in isotopically, chemically, pressure and strain tuned samples of SrTiO$_3$.

The LKSR model also predicts that the depth $\Delta\chi^{-1}(T^*)$ of the minimum should scale as $\chi^{-1}(0)$, which is partly supported from a comparison of the main plot and lower inset of Fig. \ref{fig:STO linear plots}. More importantly the polarization-acoustic phonon coupling model \citep{Khmelnitskii1971,Palova2009,Rowley2014} predicts that $\chi^{-1}(T)$ should vary as $(-T^{4})$ well below the inverse susceptibility minimum, which is in sharp disagreement with the negative quasi-linear dependence observed down to the lowest temperatures investigated \citep{Coak_Thesis_2017}. The breakdown of the distinctive $(-T^{4})$ prediction of the model is particularly striking at high pressures where temperature ranges up to two orders of magnitude below $T^*$ can readily be accessed - and the sytematic variation of the slope and extent of the quasi-linear term with $p - p_c$ is suggestive of an intrinsic phenomenon in some way connected with the quantum critical point.

This dramatic departure from the prediction of the LKSR model leads us to consider alternative explanations for the susceptibility minimum. One such alternative explanation involves the combined effects of long-range dipolar interactions between elementary
dipoles and the short range coupling of the polarization modes \citep{Conduit2010} (the mode-mode coupling) that can be represented in terms of an effective Euclidean action in a quantum description. It was suggested that this can lead to a susceptibility minimum qualitatively as predicted in the polarization-phonon coupling model, but crucially with a $(-T)$ rather than $(-T^{4})$ temperature dependence of $\chi^{-1}(T)$ below $T^*$, qualitatively as observed.

However, on closer examination we find that this negative $T$-linear form only applies to a material such as SrTiO$_3$ at relatively high temperatures, indeed at scales above that of the longitudinal polar optical frequencies, which are far above the observed $T^*$ in our experiments. In the temperature range below of the order of 10~K the dipole-dipole interaction model predicts an exponentially weak rather than a $(-T)$ temperature dependence of $\chi^{-1}(T)$, which is in sharp disagreement with observation. For this and other reasons the dipole-dipole interaction model appears to be untenable at least for the case of SrTiO$_3$ and does not explain the pressure dependence of $T^*$ and of the depth of the minimum. It is also unlikely to operate in other materials where the minimum has been observed, such as TSCC, which have ultra-weak, nearly neutral, dipoles \citep{Rowley2015}. We note, however, that the dipole-dipole interaction in polar doped alkali halides, for example, can promote antiparallel alignment of dipoles at low temperatures. This does indeed lead to a downturn in the dielectric susceptibility with decreasing temperatures at sufficiently low temperatures in these order-disorder paraelectrics that differ strongly from the displacive paraelectrics being considered here (see, e.g., \citep{Lawless1966}).

Another alternative explanation involves a possible refinement of the LKSR model , which, as already noted, predicts correctly the linear relationship between $(T^*)^2$ and $\chi^{-1}(0)$. The chief weakness of this model, namely the predicted $(-T^{4})$ temperature dependence of $\chi^{-1}(T)$ below $T^*$, compared with the observed $(-T)$ form, might be corrected via the inclusion of a low density of quasi-static modes of the lattice that can be treated effectively by classical statistics. To account for the observed $(-T)$ temperature dependence, the concentration of such modes need only be minute (below parts per million) since the Debye temperature is much larger than $T^*$, and normally outside the detection range of most probes. For example, the contribution to the specific heat capacity would be a small and virtually undetectable constant offset. Interestingly, simple numerical checks show that the inclusion of such a low density of slow classical modes, along with the acoustic phonons, leaves the pressure dependence of $T^*$, which defines a stationary point expected to be relatively insensitive to perturbations, largely unchanged. This suggests that the observed $(-T)$ variation of $\chi^{-1}(T)$ is not inconsistent with the observed linear variation of $(T^*)^2$ vs $\chi^{-1}(0)$. In contrast, however, the dependence of the depth $\Delta\chi^{-1}(T^*)$ on $\chi^{-1}(0)$ is noticeably affected by the low density of slow modes and this too is qualitatively in keeping with observation (main plot and lower inset of Fig. \ref{fig:STO linear plots}, and SI Appendix).

We now speculate on one possible origin of the proposed quasi-static modes. The LKSR model discussed thus far takes into account only the coupling of the polarization to the lattice density or to the volume strain. It has been shown that the coupling to non-uniform strain can in principle give rise to long-range strain-mediated interactions between the polarization modes (i.e., long-range mode-mode coupling). These long-range interactions are capable of producing micro-domain structures in the polarization field under certain conditions \citep{Brierley2014,Rowley2014a}, which may be expected to exhibit slow temporal fluctuations and, correspondingly, classical behaviour even at temperatures well below $T^*$. Independent evidence for the possible existence of inhomogeneities comes from a number of studies \citep{Salje2013} and, for example, from recent measurements of the thermal conductivity \citep{Martelli2018}, which reveal a surprisingly short mean-free path of phonons even in the millikelvin temperature range and in high purity single crystals of SrTiO$_3$. These speculations notwithstanding, the breakdown of the LKSR model at temperatures below the inverse suceptibilty minimum remains a mystery and potentially a major subject for future study.

We therefore conclude that the susceptibility minimum in SrTiO$_3$ can be understood largely in terms of the LKSR model. An alternative explanation for the susceptibility minimum in terms of the anharmonic effects of the long-range dipole-dipole interaction is found to be untenable at least for the case of SrTiO$_3$. Thus, we may describe the quantum paraelectric state below $T^*$ as a state in which the polarization field and the non-polar lattice field are strongly hybridized, with the emergence at still lower temperatures of a previously unknown regime characterized by a linear temperature dependence of the inverse susceptibility. This is in sharp contrast to the conventional picture in which a ferroelectric quantum phase transition separates a ferroelectric state from an unhybridized paraelectric state characterized by an activated form of the temperature dependence of the inverse susceptibility. 

We have presented new experimental findings and, in the traditional way, compared these findings with existing theoretical models.  Our analysis does not allow us to claim that the LKSR model has been `proved' even for the description of the origin of the inverse suceptibility mimimum, only that it is more realistic than other proposals.

Finally we note that a minimum of the inverse of the order parameter susceptibility is also observed in metals on the border of ferromagnetic quantum critical points.  It is possible that at least in some cases the origin of this minimum can also be attributed to the coupling of the fluctuations of the order parameter field and lattice strain or to effects of magnetostriction in place of the effects of electrostriction in the ferroelectric systems.

\section*{Methods}

High precision capacitance measurements were carried out on single crystal samples of SrTiO$_3$ from Crystal GmbH with gold electrodes vacuum evaporated onto the surfaces in a parallel-plate capacitor geometry. Measurements under hydrostatic pressure conditions were made possible by the development, in collaboration with CamCool Research Ltd, of a piston-cylinder clamp cell with miniature shielded coaxial cables running into the sample region and electrically isolated from the cell body. This eliminates stray capacitances from the wiring and allows pF capacitance signals to be measured with stabilities of up to $10^{-18}$~F, a few parts in a billion. The shield conductors of the coaxial cables were joined together at the sample position and at the measurement instrument in the standard 2-point capacitance setup. The pressure transmitting medium was Daphne Oil 7373 and pressure values, determined from the superconducting transition temperature of a tin manometer, were estimated with an accuracy of 0.5~kbar. An Andeen-Hagerling 2550A capacitance bridge was used, with an excitation amplitude voltage of 0.1~V at a fixed frequency of 1~kHz applied to the sample. The sample thickness, corresponding to capacitor plate separation, was 0.5~mm. Measurements were taken on a modified 1~K Dipper cryostat from ICE Oxford, allowing continuous stable temperature
control down to 1.2~K. Typical heating or cooling rates were held at 0.01~K per minute to allow the large thermal mass of the pressure cell to thermally equilibrate; temperature errors are of the order 10~mK at low temperature. Typical results of our measurements are shown in Figure 1 and in SI Appendix Figures S1-3.

\subsubsection*{Data availability}

All relevant data are available as a data archive from the University of Cambridge repository: \href{https://doi.org/10.17863/CAM.51389}{https://doi.org/10.17863/CAM.51389}.

\subsubsection*{Acknowledgements}

The authors would like to thank J. F. Scott, D. M. Jarvis, J. van Wezel, L. J. Spalek, S. E. Dutton, F. M. Grosche, P. A. C. Brown and C. Morice for their help in the development and evolution of this project, and D. E. Khmelnitskii, E. Baggio Saitovitch, P. Chandra, P. Coleman, V. Martelli, C. Panagopoulos, J-G. Park, G. P. Tsironis, H. Hamidov, V. Tripathi and M. Ellerby for their helpful advice and discussions. We would also like to acknowledge support from Jesus and Trinity Colleges of the University of Cambridge, the Engineering and Physical Sciences Research Council (EPSRC), the CONFAP Newton Fund, the Royal Society, IHT KAZATOMPROM, Kazakhstan and the UKRI GCRF COMPASS grant  ES/P010849/1. GGL would like to acknowledge the support of the CNPq/Science without Borders Program, CBPF, Rio de Janeiro, Brazil. Part of the work was carried out with financial support from the Ministry of Education and Science of the Russian Federation in the framework of Increase Competitiveness Program of NUST MISiS (No. K2-2017-024), implemented by a governmental decree dated 16th of March 2013, N 211.

\FloatBarrier
\newpage

\section*{Supplementary material and appendices}
\subsection*{Fitting procedures and details of analysis}

In this section we outline a collection of technical details and descriptions around values and results reported in the main text.

The uncertainty on the value of the extrapolated critical pressure, -0.7(1)~kbar, was determined from the linear fits employed, considering error bars from the dominant error in the data points. By far the largest uncertainty on each point is the pressure value - capacitance and temperature uncertainties were negligible. Changing the number of data points used had no effect on the value, within error, the data show a good linear trend. Additionally, the zero-temperature extrapolated inverse susceptibility values were found to be robust to the method chosen to establish them. At low temperature the data are almost flat; taking the values at 2 K and extrapolating to 0 K via various methods yielded very similar results. For the values reported in this work, a 2nd order polynomial was fitted to the data at very low temperature (below features of the minimum where possible), matching and preserving the shape of the data well over this limited range, and used to extrapolate to zero.

For the case of the reported characteristic temperatures shown on the phase diagram, such as $T_{QC}$ and $T_{CL}$, we stress that these are not phase transitions with well-defined temperatures, but crossovers from one regime to another where different effects are dominating the physics - no absolute temperature can be defined, we can merely be internally consistent and study their evolution. As the absolute values are arbitrary, we chose cutoffs and conditions such that fits in each region (so approaching the cutoff temperature from below and above) yield the same result at all pressure values, and to match with temperature values found from power-law analyses in Rowley et. al. (2010)\citep{RowleyThesis} and Coak (2017)\citep{Coak_Thesis_2017}. The exact criteria are given and discussed in the caption of \ref{fig:STO Tsq regime}. Temperature ranges for each fit were chosen iteratively based on the fit's valid range, and hence differ for each pressure point, see Figures \ref{fig:STO Tsq regime}-\ref{fig:STO linearFits}. Crucially, the pressure dependences of the values were seen to be insensitive to these details.

In Figure 2 of the main body of the paper $\chi^{-1}(T\rightarrow0)$ and $(T^*)^2$ are shown to vary linearly with $(p-p_{c})$. In Figure \ref{fig:STO invchiZero-vs-Tsquared} we show explicitly that $\chi^{-1}(T\rightarrow0)$ varies linearly with $(T^*)^2$.

\begin{figure}
\centering
\includegraphics[width=1\columnwidth]{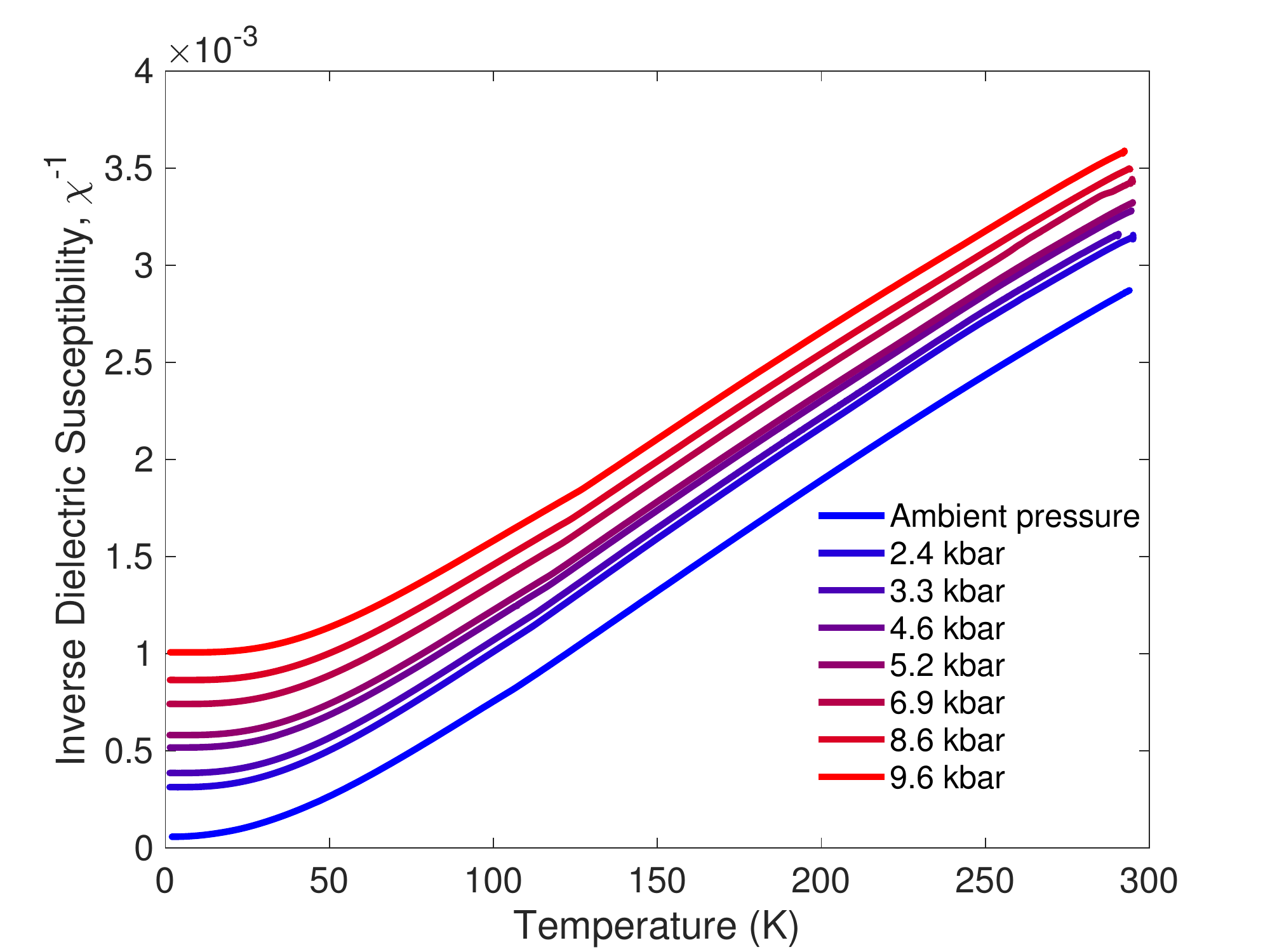}
\label{fig:STO invchi}
\caption{Inverse dielectric susceptibility $\chi^{-1}$ of SrTiO$_3$ plotted against temperature for applied pressures ranging from 0 (blue) to 9.6~(red) kbar.}
\end{figure}

\begin{figure}
\centering
\includegraphics[width=1\columnwidth]{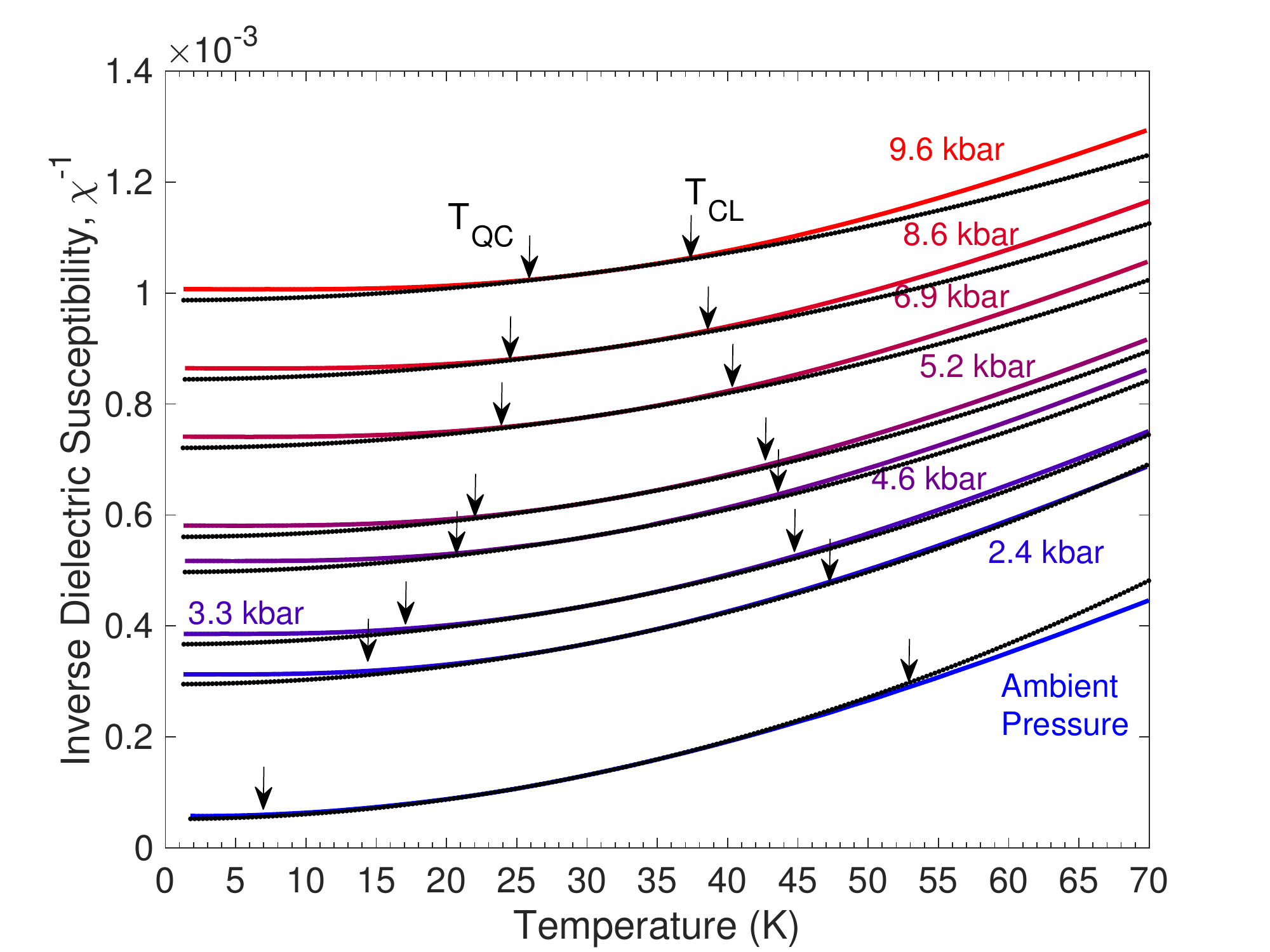}
\caption{The measured values of $\chi^{-1}$ as a function of temperature at different fixed pressures (shown as coloured dots) were fitted to an equation of the form $\chi^{-1}=A+BT^{2}$ over the temperature range 0 to 70~K ($A$ and $B$ are fitting parameters). The fitted curves are shown as black dotted lines for each pressure. The lower and upper cross-over temperatures $T_{QC}$ and $T_{CL}$, marked with arrows, were defined as occurring when the experimental value of $\chi^{-1}$ deviated from the fitted value by 1\%.}
\label{fig:STO Tsq regime}
\end{figure}

\begin{figure}
\centering
\includegraphics[width=1\columnwidth]{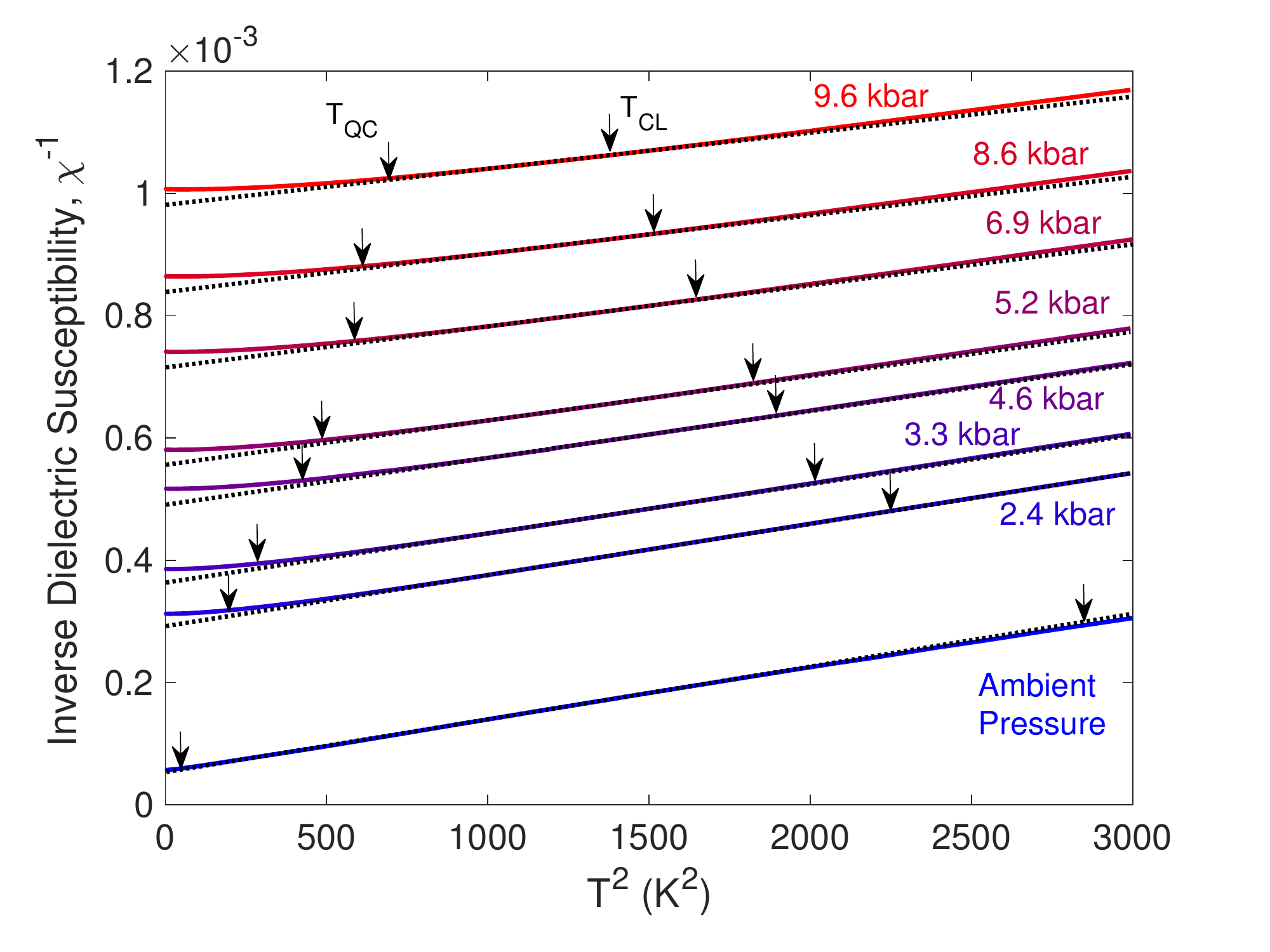}
\caption{Inverse dielectric susceptibility $\chi^{-1}$ of SrTiO$_3$ plotted against the square of temperature for applied pressures ranging from 0 (blue) to 9.6~(red) kbar. The regions where the data are linear, indicating a $T^2$ temperature dependence, lie between the characteristic crossover temperatures $T_{QC}$ and $T_{CL}$, marked with arrows. $T_{QC}$ and $T_{CL}$, described more in the main text, were determined as the temperatures at which the measured value of $\chi^{-1}$ deviated from the fitted value of $\chi^{-1}$ by 1\%.}
\label{fig:STO invchi-vs-Tsquared}
\end{figure}

\begin{figure}
\centering
\includegraphics[width=1\columnwidth]{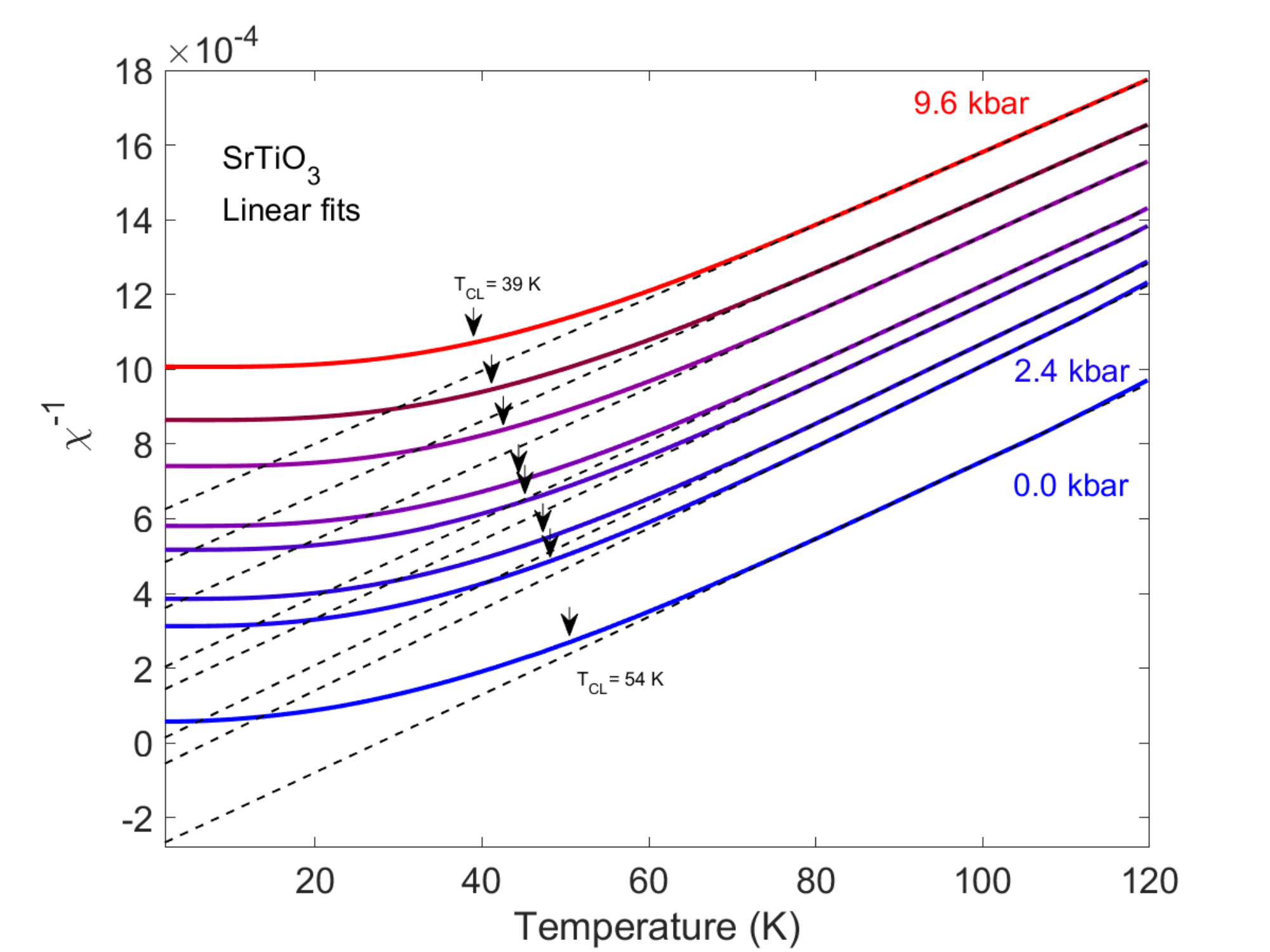}
\caption{Inverse dielectric susceptibility $\chi^{-1}$ of SrTiO$_3$ plotted against temperature for applied pressures ranging from 0 (blue) to 9.6~(red) kbar. Dashed lines show linear Curie-Weiss classical fits to the high-temperature data. The temperatures $T_{CL}$ where the temperature dependence becomes dominated by $T^2$ quantum critical behaviour (arbitrarily defined as a fixed 8\% divergence of the fits to match with other methods) are marked with arrows.}
\label{fig:STO linearFits}
\end{figure}

\begin{figure}
\centering
\includegraphics[width=1\columnwidth]{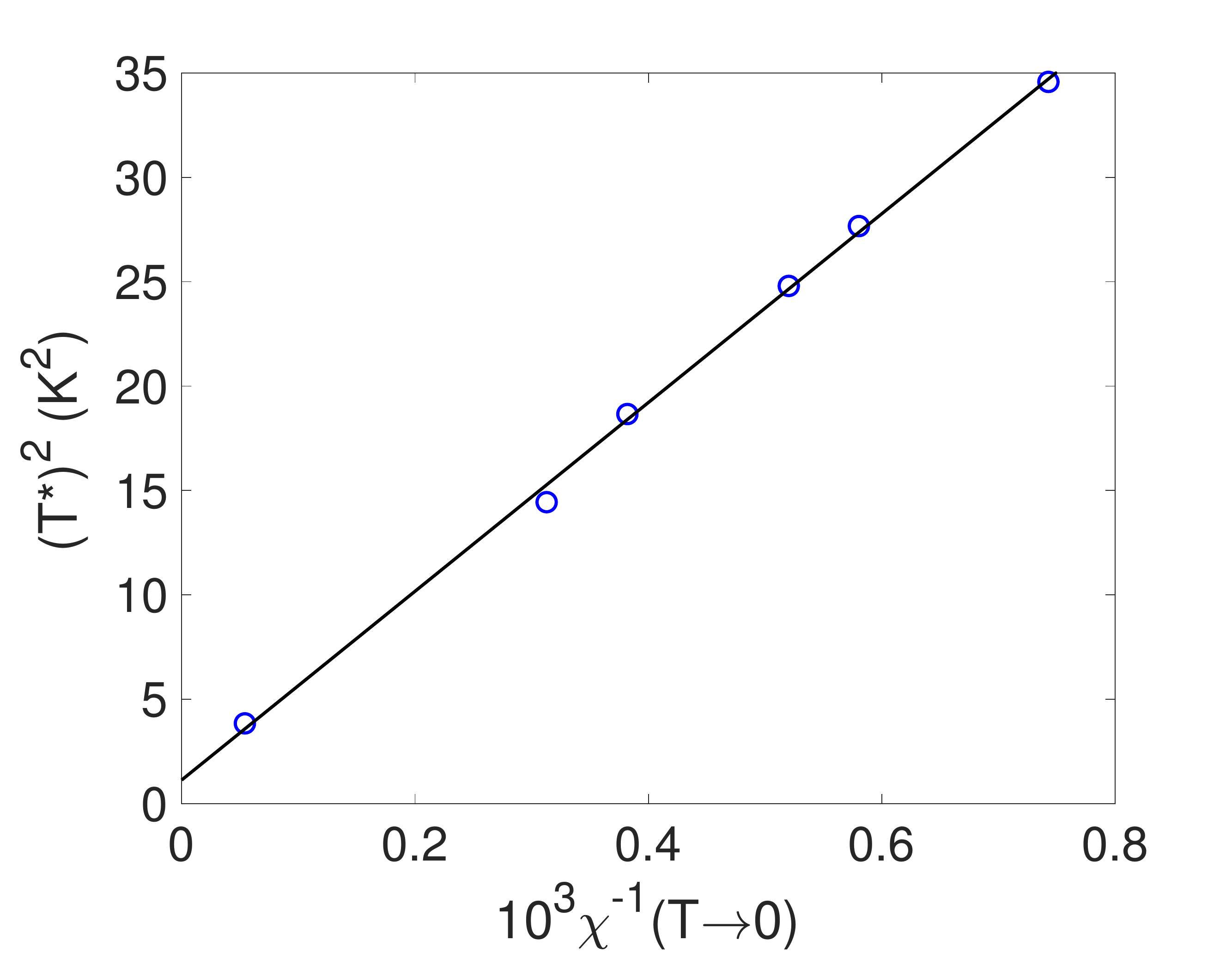}
\caption{The square of the temperature of the inverse susceptibility minimum based on the data shown in Fig. 2 of the main text, plotted against the extrapolated zero-temperature inverse susceptibility. Solid line shows a linear fit to the data.}
\label{fig:STO invchiZero-vs-Tsquared}
\end{figure}

\FloatBarrier
\newpage
\subsection*{Self-Consistent Phonon Theory}

The calculations of the dielectric susceptibility, $\chi\left(T\right)=\varepsilon_{r}(T)-1$, were carried out in terms of the self-consistent phonon model for a simple displacive ferroelectric in three dimensions, see Refs \citep{Rowley2014,RowleyThesis}
and references therein. We consider first the effects of the self interaction of the electric polarization field as represented by a
local quartic term in the effective Lagrangian density, expressed as a function of the polarization field and its temporal and spatial
gradients. In the mean field approximation for the self-interaction the polarization correlation wavevector, or inverse correlation length, $\kappa(T)$, is determined from the self-consistent equation

\begin{equation}
\begin{split}
\kappa^{2}(T)=&\kappa^{2}(0)+\zeta\int dq~q^{2}\left[\sum_{n}\frac{T}{\kappa^{2}(T)+q^{2}+\omega_{n}^{2}}\right. \\
&\left.-\frac{1}{2\pi}\int\frac{d\omega}{\kappa^{2}(0)+q^{2}+\omega^{2}}\right]\label{eq:1}
\end{split}
\end{equation}
where the wavevectors $q$, $\kappa(T)$ and $\kappa(0)=\kappa(T=0)$ are in units of the relevant Debye wavevector $\Lambda$, $T$ and are $\omega$ are in units of the relevant Debye temperature $\theta$ and $\omega_{n}=2\pi nT$, where $n$ is an integer. The cut-offs in wavevector and frequency are taken to be unity in these units (i.e., $0<q<1$ and $-1<\omega<1$). The dimensionless correlation wavevector $\kappa(T)$ and the dimensionless coupling parameter $\zeta$ are given, in terms of the model parameters defined in the next section and in Ref \citep{Rowley2014}, by

\begin{equation}
\kappa^{2}(T)=\frac{\Delta^{2}\chi(0)}{\upsilon^{2}\Lambda^{2}\chi(T)}\label{eq:2}
\end{equation}
and
\begin{equation}
\zeta=\frac{5\varepsilon_{0}\hbar b\Delta^{4}\chi^{2}(0)}{3\pi^{2}\upsilon^{3}}\label{eq:3}
\end{equation}
where $\chi\left(0\right)=\chi(T=0)$, $\Delta$ and $\upsilon$ define the $"T=0"$ spectrum of the critical transverse optical modes and $b$ is the coefficient of the anharmonic term in the equation of state assumed to have the analytic form $\varepsilon_{0}E=P/\chi(T)+bP^{3}$, where $E$ is the electric field that stabilizes the polarization $P$. Thus $\kappa^{2}(T)$ is a dimensionless measure of the inverse dielectric susceptibility. We also note that the square of the gap of the spectrum of critical transverse optical modes is expected to be proportional to the inverse susceptibility, so that $\Delta^{2}(T)=\chi^{-1}(T)\Delta^{2}/\chi^{-1}(0)$, where $\Delta=\Delta(0)$.

The coupling between the critical transverse optic mode and the acoustic mode, or more precisely the volume strain, leads to a further correction to $\kappa^{2}(T)$ of the form 
\begin{equation}
\begin{split}
\delta\kappa^{2}(T)=&-\lambda\int dq~q^{4}\left[\sum_{n}\frac{T}{(\kappa^{2}(T)+q^{2}+\omega_{n}^{2})(q^{2}+\eta^{2}\omega_{n}^{2})}\right. \\
&\left.-\frac{1}{2\pi}\int\frac{d\omega}{(\kappa^{2}(0)+q^{2}+\omega^{2})(q^{2}+\eta^{2}\omega_{n}^{2})}\right]\label{eq:4}
\end{split}
\end{equation}
where $\eta$ is the ratio of the velocity of critical transverse optical modes to the velocity of the acoustic mode and $\lambda$
is a dimensionless electrostrictive coupling constant. We have estimated the latter via the strain dependence of the energy gap of the order parameter mode \citep{Rowley2014} that suggests 
\begin{equation}
\lambda\thickapprox\frac{3K\zeta}{20\varepsilon_{0}bp_{c}^{2}\chi^{2}(0)}\label{eq:5}
\end{equation}
where $\zeta$ is the dimensionless parameter defined in Eq. \ref{eq:3}, $K$ is the bulk modulus and $p_{c}$ is the critical pressure (negative for SrTiO$_3$) where the $T=0$~K correlation wavevector is expected to vanish. We note that the electrostrictive effect is also expected to make a negative contribution to the mode-coupling parameter, $b$, and lead potentially to a first order transition. Numerical estimates suggest, however, that the transition should remain second order under our experimental conditions in agreement with measurements down to $0.3$~K in SrTiO$_3$ \citep{Rowley2014}. Nonetheless, as one approaches even closer to the quantum critical point than explored in our experiments, a sign change in $b$ may occur.

\begin{figure}
\centering
\includegraphics[width=1\columnwidth]{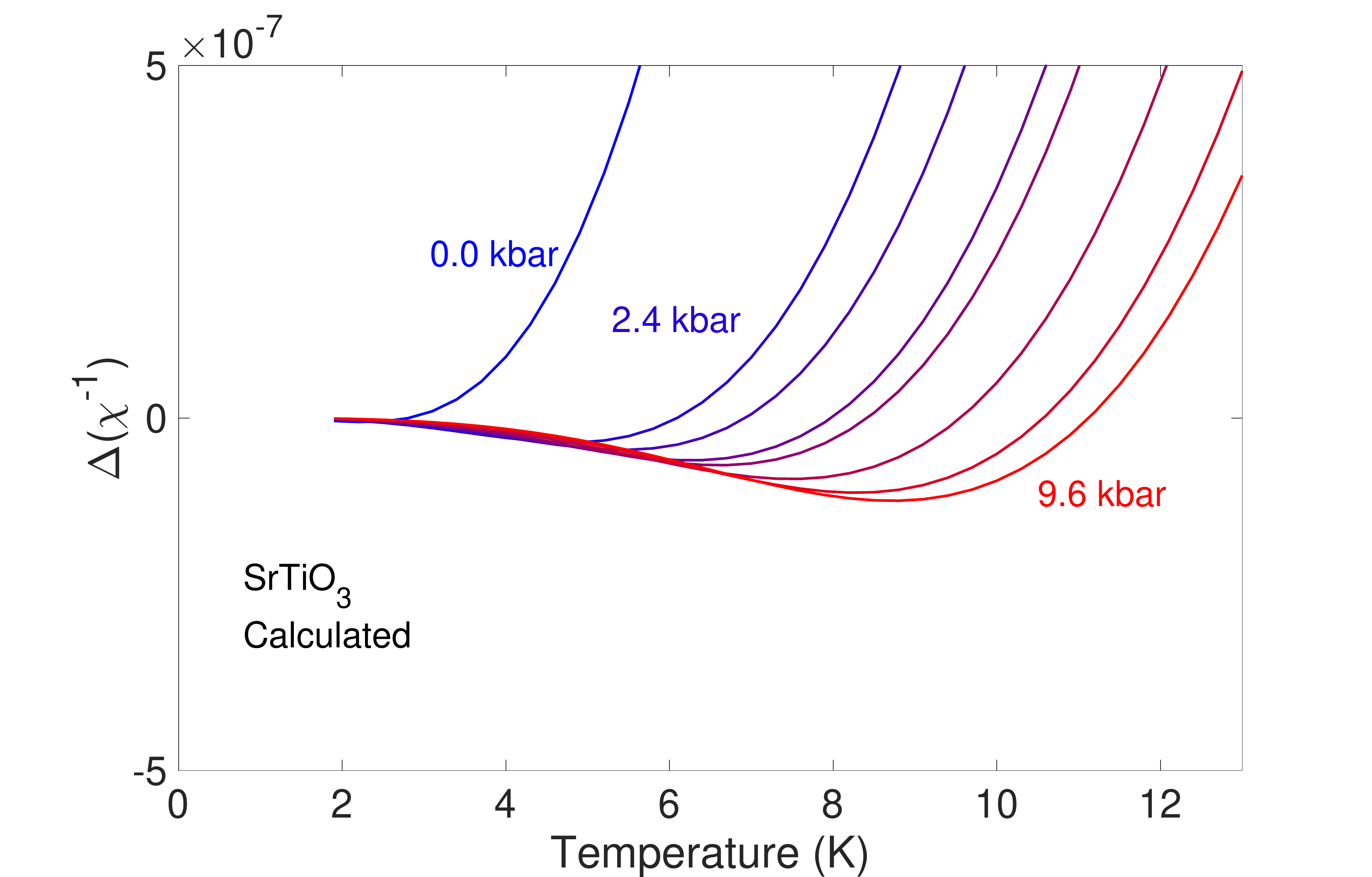}
\includegraphics[width=1\columnwidth]{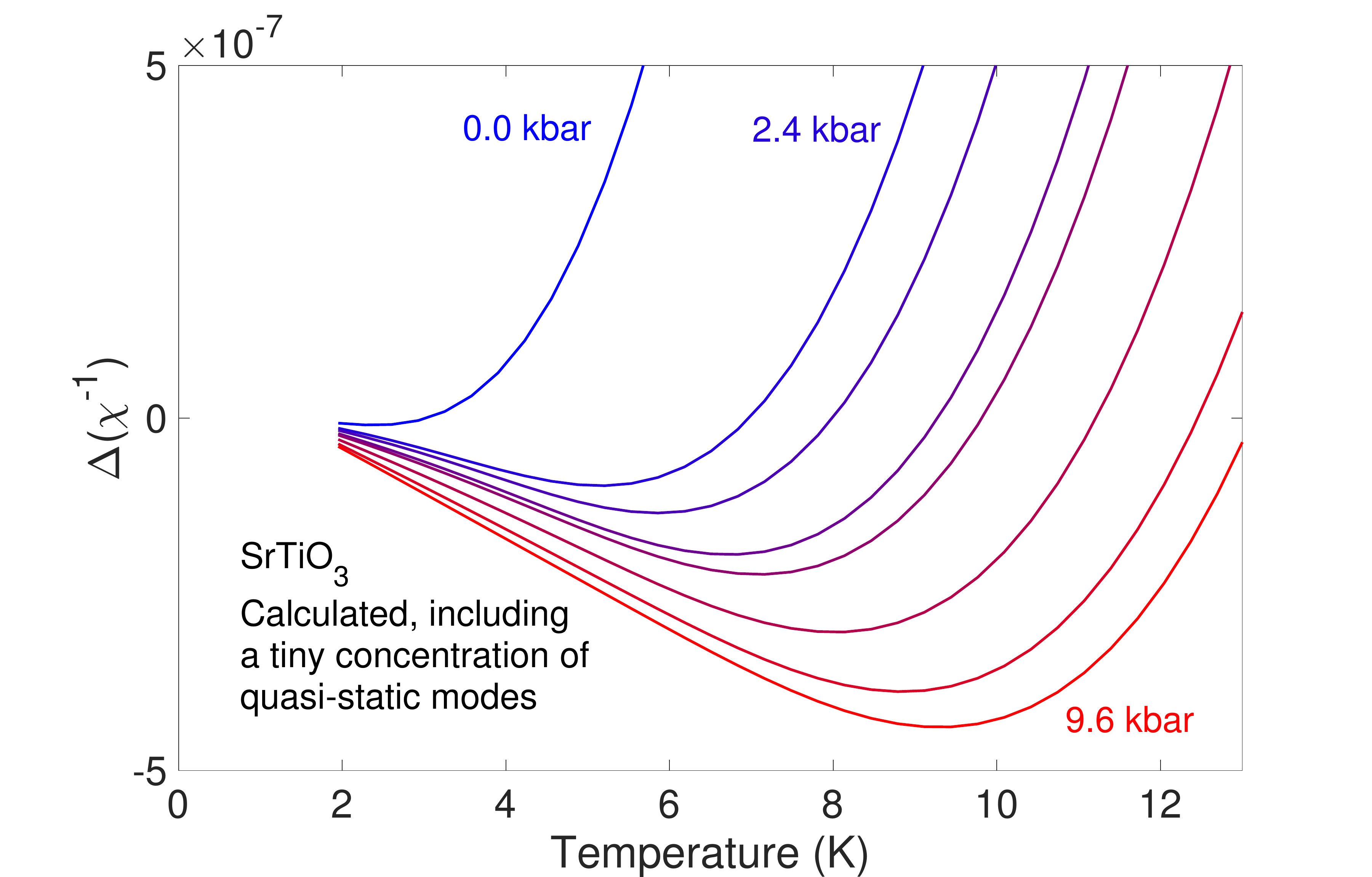}
\caption{Upper: calculated temperature variation of the inverse susceptibility at different applied pressures. Predictions of the self-consistent phonon model, including electrostrictive coupling, i.e. of the LKSR model, for the temperature dependence of $\Delta\chi^{-1}(T) = \chi^{-1}(T)-\chi^{-1}(T_{ref})$, with $T_{ref}$ = 1.3~K. Lower: the same calculation, but including a small concentration of quasi-static modes, as discussed in the main text.}
\label{fig:Calculated-pressure-dependence}
\end{figure}

We note that in the limit where $\kappa^{2}$ is comparable to or greater than $q^{2}$ in the above equations (for all $q<\Lambda$), the above model in the absence of the electrostrictive effect and self-consistency reduces to the Barrett model. The latter cannot describe the quantum critical regime where the frequencies of the modes are strongly wavevector and temperature dependent, nor below the crossover temperature $T^*$ in general where the electrostrictive effect changes the qualitative nature of the quantum paraelectric state. 

The prediction of the above model (defined by Equations \ref{eq:1}-\ref{eq:5}) for the temperature dependence of the susceptibility at a series of applied hydrostatic pressures for model parameters relevant to SrTiO$_3$ (next section) is shown in Figure 4 in the main body of the paper and in Figure \ref{fig:Calculated-pressure-dependence}.
For calculations at finite pressures the ambient pressure parameters $\chi^{-1}(0)$ and $\Delta$ are replaced, respectively, by $\chi^{-1}(0)(1-p/p_{c})$ and $\Delta(1-p/p_{c})^{1/2}$, while $b$ and $v$ are held constant. The results exhibit a minimum arising from the electrostrictive coupling with the minimum temperature $T^*$ and depth $\Delta\chi^{-1}(T^{*})$
that decrease with decreasing pressure and vanish at the ferroelectric quantum critical point at $p_{c}=-0.7$~kbar. As shown in Figures 4 and \ref{fig:Calculated-pressure-dependence} the model predicts that $(T^*)^2$ and the depth $\Delta\chi^{-1}(T^{*})$ should both be proportional to $(p-p_{c})$. The former prediction, in particular, is in excellent agreement with experiment (c.f. Figure 4 and upper inset of Figure 2 in the main body of the paper). However, the electrostrictive coupling model also predicts a $(-T^{4})$ temperature dependence of $\Delta\chi^{-1}(T^{*})$ well below $T^*$, which is inconsistent with the negative quasi-linear variation observed (cf. Figures 1 and 4 in the main body of the paper). 

In an effort to understand this unexpected behaviour, we have also considered the possible roles of the long-range dipole-dipole interaction and of quasi-static inhomogeneities of the lattice or micro domains predicted to arise from anisotropic couplings (including wavevector dependence) of the strain and electric polarization fields. The main effect of the dipolar interaction is to modify a degeneracy factor defining the dimensionless self-interaction coupling parameter already included in the expression for $\zeta$ given in Equation (\ref{eq:3}). Additional predicted consequences of the dipolar interaction are found not to be relevant to SrTiO$_3$ at low temperatures as discussed in the main body of the paper. 

Finally we consider more closely the temperature dependence of the electrostrictive correction. Via contour integration Equation (\ref{eq:4}) can be rewritten in the form (Palova et al.\citep{Palova2009})
\begin{equation}
\delta\kappa^{2}(T)\propto-\lambda\sum_{q}\frac{q^2}{\omega_{mq}\omega_{\phi q}}\{\omega_{mq}n(\omega_{\phi q}/T)-\omega_{\phi q}n(\omega_{m q}/T)\}\label{eq:6}
\end{equation}
where $\omega_{mq}$ and  $\omega_{\phi q}$  represent the spectra of the order parameter modes and the acoustic modes, respectively, and $n(\omega/T)$ is the Bose function. For $T$ well below $T^*$
\begin{equation}
\delta\kappa^{2}(T)\propto-\lambda\zeta\sum_{q}\frac{q^2}{\omega^{2}_{mq}\omega_{\phi q}}n(\omega_{\phi q}/T)\propto-T^{4}\label{eq:7}
\end{equation}
i.e., $\delta\kappa^{2}(T)$ is expected to fall as the fourth power of $T$ or as the thermal energy of the acoustic modes in three dimensions. Equation (\ref{eq:7}) suggests that if in place of the acoustic modes we consider quasi-static modes with a characteristic energy scale smaller than $T$, such that the Bose function reduces to $T/\omega$, a negative linear rather than quartic shift in T would be expected to arise. Because of the high energy scale (i.e., the Debye energy) of the acoustic modes, even a very tiny concentration of quasi-static modes postulated here can produce a dominant contribution in the low temperature range of interest well below $T^*$ (see main body of paper).

\subsection*{Model Parameters}

The starting self-consistent phonon model for a displacive ferroelectric is defined mainly in terms of four temperature independent parameters, which we take to be $\chi^{-1}(0)$ , $b$, $\Delta$, and $v$. The parameters $\chi^{-1}(0)$ and $b$ were obtained from the intercept and slope, respectively, of plots of $\varepsilon_{0}E/P~vs~P^{2}$ at 0.3~K \citep{RowleyThesis}, where $E$ is the electric field (0 to 15~kV/cm in our measurements) and $P$ is the electric polarization,
given an equation of state of the analytic form $\varepsilon_{0}E/P=\chi^{-1}(0)+bP^{2}$. The parameters $v$ and $\Delta$ were determined by comparing the data from inelastic neutron and Raman scattering experiments at 4 K to an equation of the form $\omega_{q}^{2}=\Delta^{2}+v^{2}q^{2}$, which defines the low $q$ and low $T$ dispersion in the paraelectric
state of the transverse-optic phonons that are the soft modes of the incipient ferroelectric state (see Ref \citep{Rowley2014} and references therein). Values for $\chi^{-1}(0)$ , $b$, and $v$ for SrTiO$_3$ in the low temperature limit used in the calculations were set equal to $2\times10^{4}(=\varepsilon_{r}(0)-1)$, $0.07~C^{2}m^{4}$ and $8100~ms^{-1}$, respectively \citep{Rowley2014}. The results of calculations presented are based on three values of
the gap $\hbar\Delta/k_{B}$ equal to 6, 12 and 24~K. This range covers the range of values of the low temperature gap estimated via other experiments (see, e.g., Yamada and Shirane \citep{Yamada1969}).

We note that the square of the gap is expected to be proportional to the inverse susceptibility, so that $\Delta^{2}(T)=\chi^{-1}(T)\Delta^{2}/\chi^{-1}(0)$, where $\Delta=\Delta(0)$, and that $v$ reduces to the speed of sound of the critical transverse optical modes when the gap vanishes. The relatively small value of $\Delta$ implies that SrTiO$_3$ is close to a displacive ferroelectric quantum critical point at ambient pressure. For a Debye wavevector $\Lambda$ set equal to $\pi$ divided by the lattice constant (0.39 nm), the dimensionless self-interaction coupling constant $\zeta$ defined in Equation (\ref{eq:3}) is approximately 0.8.

Also we have taken $\eta=2$, $\theta=600$~K, $\Delta=12$~K, $p_{c}=-0.7$~kbar and $K=180$~GPa, so that the dimensionless electrostrictive coupling constant $\lambda$ defined in Equation (\ref{eq:5}) is approximately 0.03 (see Ref \citep{Rowley2014} and references therein). For calculations at finite pressures $\chi^{-1}(0)$ and $\Delta$ are replaced, respectively, by $\chi^{-1}(0)(1-p/p_{c})$ and $\Delta(1-p/p_{c})^{1/2}$, while $b$ and $v$ are held constant. 

\end{document}